\documentclass{rmaa}[journal]
\usepackage{amsmath}
\usepackage{xcolor}
\usepackage[normalem]{ulem}

\title{Collisions in a system of conical jet/counterjet outflows}
\author{
  A. C. Raga\altaffilmark{1}, Z. Meliani\altaffilmark{2}, A. Rodr\'\i guez-Gonz\'alez\altaffilmark{1},
  S. Cabrit\altaffilmark{3}, G. Pineau des For\^ets\altaffilmark{3}, J.I. Castorena, \altaffilmark{1},A. Esquivel\altaffilmark{1}}
\altaffiltext{1}{Instituto de Ciencias Nucleares, UNAM, M\'exico}
\altaffiltext{2}{LUTH, Observatoire de Paris, CNRS, PSL, Universit\'{e} de Paris; 5 Place Jules Janssen, 92190 Meudon, France}
   \altaffiltext{3}{Observatoire de Paris, PSL University, Sorbonne University, CNRS, LERMA, 75014 Paris, France}

\fulladdresses{
\item A. Rodríguez-González: Instituto de Ciencias Nucleares,
Universidad Nacional Aut\'onoma de M\'exico, Ap. 70-543, 04510 CDMX,
M\'exico (ary@nucleares.unam.mx)
}

\shortauthor{Raga et al.}
\shorttitle{Collisions of conical outflows}
\ReceivedDate{\today} 
\AcceptedDate{Year Month Day} 

\resumen{Las estrellas se forman, predominantemente, en cúmulos compactos. Los flujos de gas eyectados por proto-estrellas pueden cruzarse e interactuar entre sí, lo que resulta en una compleja interacción. Estas interacciones afectan la dinámica, morfología y evolución de estos flujos. La determinación de la probabilidad de encuentro entre ellos requiere un enfoque bayesiano que considere la colimación, longitud (o edad), la separación entre objetos estelares jóvenes en cúmulos. En este estudio, empleamos un enfoque de Monte Carlo para estimar esta probabilidad como función del ángulo de apertura del chorro y la relación entre la longitud del chorro y la separación entre estrellas. Proponemos una función que predice el número de interacciones dentro de un cúmulo basada en el ángulo de apertura de los flujos de gas eyectados por protoestrellas. }

\abstract{Stars predominantly form in compact, non-hierarchical clusters. The gas outflows ejected by protostars can intersect and interact with each other, resulting in complex interactions that affect the dynamics, morphology, and evolution of these outflows. Determining the probability of an encounter between them requires a Bayesian approach that considers the collimation, length (or age), and separation between young stellar objects in the clusters. In this study, we employ a Monte Carlo approach to estimate this probability as a function of the jet opening angle and the ratio between the jet length and the separation between stars. We propose a function that predicts the number of interactions within a cluster based on the opening angle of the gas outflows ejected by protostars.}


\begin{document}
\maketitle

\section{Introduction}
The interaction of molecular outflows is a fascinating phenomenon for understanding the early evolution of low-mass stellar objects in star-forming regions. It has been observed that low-mass objects form through an accretion process, resulting in the production of collimated outflows that eject material from very close to the formed object, known as astrophysical jets.

In \citep{nony2020}, the mass structure of the complex W43 cloud was studied, focusing on pre-stellar and protostellar objects in this mini-starburst. They found that 51 out of 127 cores in the cloud are associated with molecular outflows, while some lack associated outflows. These numerous objects are located within a filamentary structure approximately 1 pc in length along its major axis. The high probability of collision and interaction between the molecular outflows in this region is evident from ALMA telescope images in CO (2-1) (see Fig. 2 in \citealt{nony2020}).

Similar interactions between molecular outflows had been previously reported in association with the object BHR71, which exhibits a pair of outflows emanating from two young protostars within a Bok's globule. The outflows in BHR 71 interact with the surrounding material in the molecular cloud, causing shocks and gas compression. These molecular outflows are likely colliding, as indicated by an increase in the brightness of the CO emission, a dispersion of velocities in the impact zone, and a change in the orientation of one of the outflows (IRS 2), as shown by \citet{zapata2018}.

More recently, in Cep E-south, a cavity has been observed in the envelope of the main jet due to a perpendicular secondary jet interacting with it, destroying molecules primarily in CO (2-1) emission (see \citealt{lefloch2015}). Furthermore, \citet{rg2025} demonstrated that this interaction of molecular outflows drastically changes the structure of the envelopes associated with jets from Class 0/I objects.

In this study, we aim to estimate the probability of such interactions occurring and how this is related to their separation distance and opening angle. The paper is organized as follows: In Section 2, we calculate the probability of interaction between two conical flows as a function of their opening angles and the distances each travels. An extension to a system with more than one jet is explored in Section 3. We make theoretical assumptions to test our approximations, detailed in Section 4. The experiments based on gas dynamics are described in Section 5. Section 6 is dedicated to applying our models to predict interacting conical flows. Finally, we present the conclusion of our work in the last chapter.
\begin{figure}[!t]
\includegraphics[width=\columnwidth]{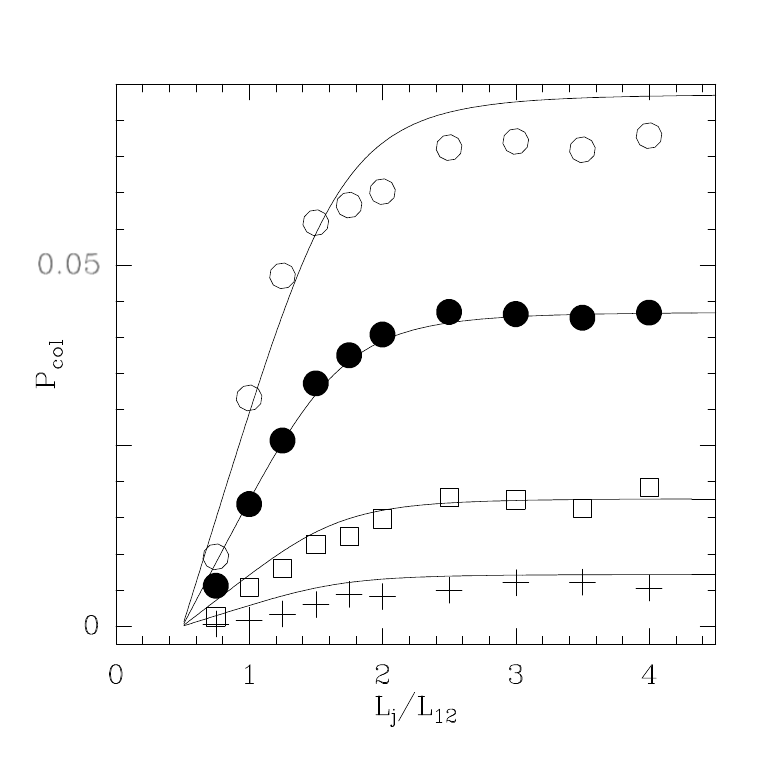}
\caption{The probability of collision between two ``double cone'' outflows
  as a function of the $L_j/L_{12}$ jet length to source separation ratio.
  The results of sets of $10^4$ ``Bernoulli experiments'' for double cones
  with half-opening angles $\alpha=2.5^\circ$ (crosses), $5^\circ$ (squares),
  $10^\circ$ (full circles) and $15^\circ$ (open circles) are shown.
  The solid curves are obtained from the analytic fit of equation (\ref{p})
  for the same $\alpha$ values.}
\label{fig1}
\end{figure}

\begin{figure}[!t]
\includegraphics[width=\columnwidth]{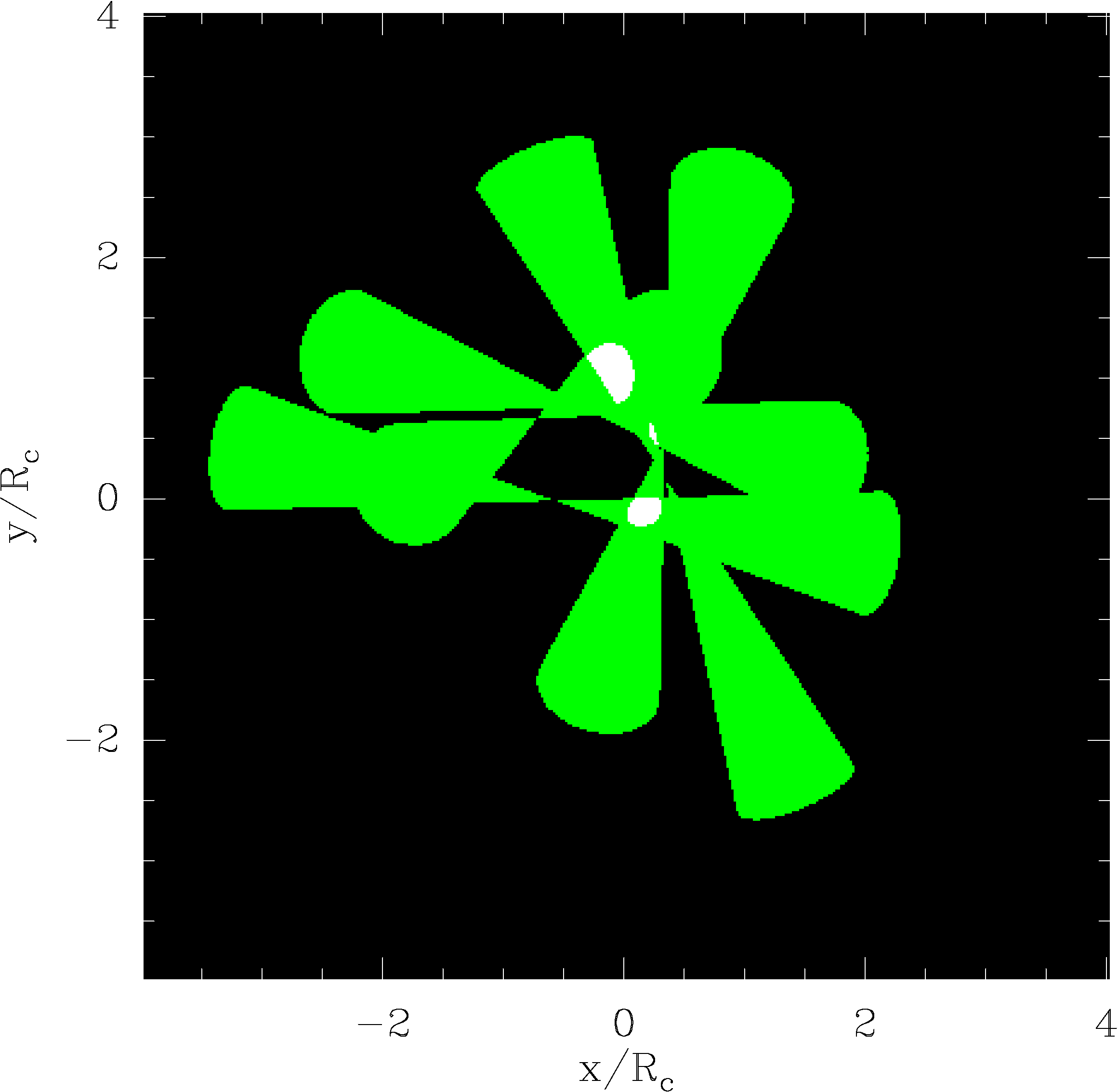}
\caption{Configuration of a system of $N_j=5$ biconical bipolar
  outflows with a half-opening angle $\alpha=10^\circ$. The outflow
  directions are randomly chosen, and the source positions are obtained
  by sampling a uniform source distribution within a sphere of radius
  $R_c$. The projection of the cones on the $xy$-plane is shown in green,
  and the regions of physical (3D) superpositions between cones are
  shown in white. The outflow cones have $L_j=3R_c$ lengths.}
\label{fig2}
\end{figure}

\begin{figure}[!t]
\includegraphics[width=\columnwidth]{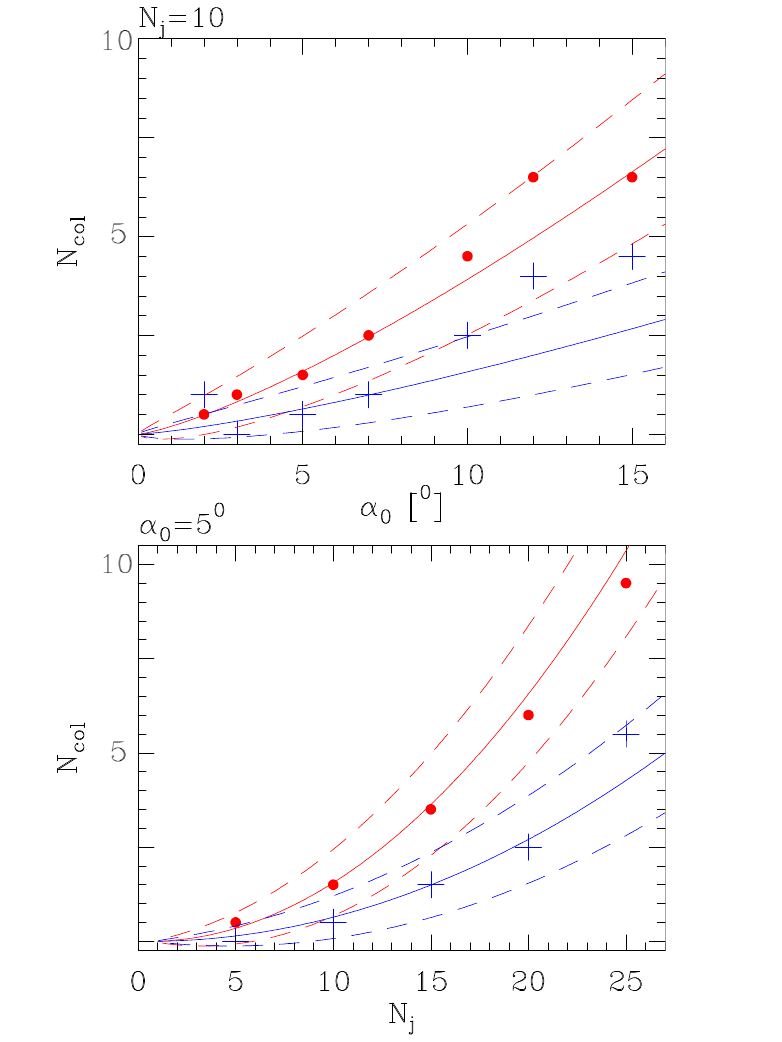}
\caption{Top: Number of jet superpositions as a function of the half-opening
  angle $\alpha$ for a system of $N_j=10$ bi-conical outflows. The results
  for (single) Bernoulli experiments of randomly directed jets from sources
  with a uniform spatial distribution within a sphere of radius $R_c$ have
  been computed for cones of length $R_j=R_c$ (blue crosses) and $R_j=3R_c$
  (red circles). The solid curves correspond to the results obtained
  with the analytic fit of equation (\ref{nc}), and the dashed curves are
  the corresponding $\pm \sigma$ envelopes. Bottom: number of jet
  superpositions for a system of cones with $\alpha=5^\circ$ as a function
  of the number $N_j$ of jets with sources within a sphere of radius $R_c$.
  The results for jets of lengths $R_j=R_c$ (blue crosses and curves)
  and $R_j=3R_c$ (red circles and curves) are shown.}
\label{fig3}
\end{figure}

\begin{figure}[!t]
\centering
\includegraphics[width=0.5\columnwidth]{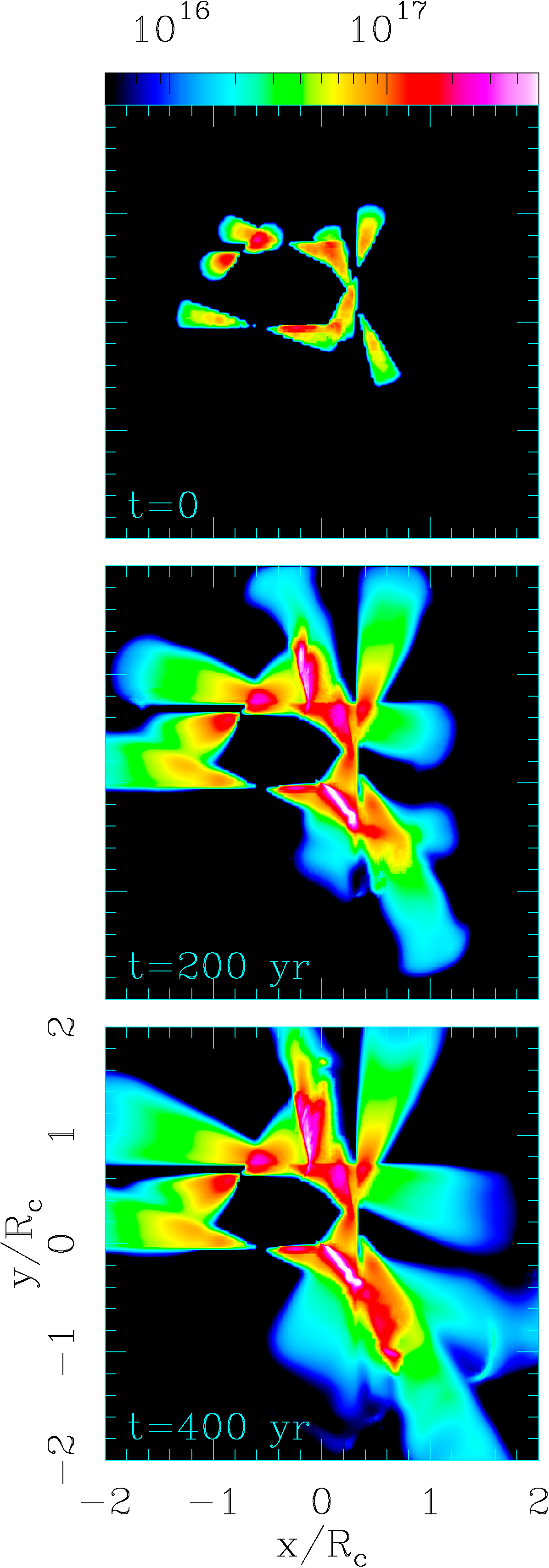}
\caption{Column density time frames for $t=0$ (the initial condition,
  top frame), 200 (center) and 400~yr time-integration (bottom) obtained
  from the 3D gasdynamic simulation described in the text. The column
  densities are displayed with the logarithmic scale given (in cm$^{-2}$
  by the top bar. The column densities correspond to integrations along
  the $z$-axis, and the $(x,y)$ coordinates are given in units of
  the $R_c=5\times 10^{16}$~cm radius of the outflow source position
  distribution. The source position and direction values for the
  5 bipolar outflows are the same as the ones of the geometrical model
  of Figure 2.}
\label{fig4}
\end{figure}

\section{Interaction between two outflows}

We first consider a problem of two jet/counterjet, conical outflows
of half-opening angle $\alpha$ and length $L_j$ (measured from the source
to the tip of the jet or the counterjet), with a separation $L_{12}$
between the two sources. To evaluate the probability $p$ of
having a physical collision between the two outflows, we carry out
a Montecarlo simulation of $N$ ``Bernoulli experiments''
in which the orientations
of the two outflows are chosen randomly, and we count the number
$N_c$ of cases which lead to a collision between the lobes of the
two outflows. The simulation then gives an estimate $P_{col}=N_{col}/N$ for
the two-outflow collision probability.

The collision probability $P_{col}$ as a function of the jet length to
source separation ratio $L_j/L_{12}$ (obtained from simulations with
$N=10^4$ experiments) is shown in Figure 1. The probability
is given for four values of the half-opening angle
of the outflow cones: $\alpha=2.5$, 5, 10 and $15^\circ$.

As expected, the collision probability $P_{col}$ is zero for $L_j/L_{12}\leq 1/2$
(since this condition implies that the outflows never touch). For
$L_j/L_{12}>1/2$, $p$ first grows and then reaches an approximately
constant limit  $L_j/L_{12}>2.5$ (see Figure 1).

Also, in Figure 1, it is clear that the collision probability $p$ is
a growing function of the half-opening angle $\alpha$. We propose
a form $P_{col} \propto \alpha^q$. For small opening angles:
\begin{itemize}
\item in perpendicular jet collisions, the solid angle subtended
  by the obstructing jet is $\propto \alpha$, so we would expect
  $q\approx 1$,
\item in collisions of closely aligned jets, the solid angle
  of the obstructing jet is $\propto \alpha^2$, giving $q\approx 2$.
\end{itemize}
We find that a $q=1.3$ value approximately fits the angular dependence
of the collision probability obtained from our montecarlo simulations.

We propose a fitting formula of the form:
$$P_{col}(L_j/L_{12},\alpha)=$$
\begin{equation}
    \left\{\left[0.07\left(\frac{L_j}{L_{12}}-\frac{1}{2}\right)\right]^{-4}+
  17500\right\}^{-1/4}\left(\frac{\alpha}{10^\circ}\right)^{1.3}\,,
  \label{p}
\end{equation}
which has an initially linear growth for $L_j/L_{12}>1/2$, and
goes to a constant value for $L_j/L_{12}>>1$. The $p$ vs.
$L_j/L_{12}$ curves obtained for $\alpha=2.5$, 5, 10 and $15^\circ$
are shown in Figure 1, showing a reasonable agreement with the
results of the Montecarlo simulations.

\section{Collisions in a system of $N_j>2$ outflows}

We now consider a system of $N_j$ identical biconical jet/counterjet
systems with the same half-opening angle $\alpha$ and length $L_j$
(from the source to the end of the cones). We assume that the outflow sources
(at the tips of the cone pairs) are uniformly distributed within
a sphere of radius $R_c$.

For a uniform, spherical distribution of sources with radius
$R_c$, the average separation between source pairs is $\approx 1.03\,R_c$
and the root mean square separation is $\approx 1.09\,R_c$. We therefore
assume that the average probability $p_c$ for a collision between
jet pairs is given by equation (\ref{p}) with $L_{12}=R_c$. This
approximation of taking the same probability for the interactions
of all outflow pairs is, of course, correct for the case
of $L_j/R_c\gg 1$, for which the interaction probability does not
depend on the separation between the outflow sources.

We now consider a set of Bernoulli experiments consisting
of the success/failure events (with success probability $P_{col}$, see
above) of the interaction between a given
outflow $i$ (with $i=1\to N_j$) with an outflow $j$ (with $j$ having
the $N_j-1$ values of the remaining sources). If the processes
follow a binomial distribution, the expected number of collisions is
\begin{equation}
  N_{col}=(N_j-1)N_jP_{col}/2\,,
  \label{nc0}
\end{equation}
with a standard deviation
$\sigma=\sqrt{(1-P_{col})N_{col}}\approx \sqrt{N_{col}}$ (the
second equality being appropriate for $P_{col}\ll 1$).

Equation (\ref{nc}) is obtained for a binomial statistical distribution.
When $N_c$ approaches $N_j$, a more complex hypergeometric
distribution  should be considered. We do not study this limit because
a real system of outflows with many jet interactions defies our simple
picture of a system of biconical, geometrically fixed structures!

Now, combining equations (\ref{p}) (with $L_{12}=R_c$) and
(\ref{nc0}) we obtain:
$$N_{col}=(N_j-1)N_j\times$$
\begin{equation}
\left\{\left[0.07\left(\frac{L_j}{R_c}-
    \frac{1}{2}\right)\right]^{-4}+
  17500\right\}^{-1/4}\left(\frac{\alpha}{10^\circ}\right)^{1.3}/2\,.
  \label{nc}
\end{equation}
This is the expected number of jet collisions for a system of $N_j$
conical bipolar outflows of half-opening angle $\alpha$
and length $L_j$ from sources distributed within a radius sphere $R_c$. The standard deviation of the number of collisions
with respect to this expected value is $\sigma\approx \sqrt{N_{col}}$.

\section{Numerical experiments of systems of many bi-conical outflows}

We now carry out experiments in which we randomly place $N_j$ outflow sources 
within a sphere of radius $R_c$, also choosing a random outflow
direction for each source. We then place biconical outflow structures
of half-opening angle $\alpha$ and length $L_j$ (for each of the cones)
at the randomly chosen positions and directions. Finally,
in the resulting structure, we count the regions of superposition
of the outflow cones.

An example of a system with $N_j=5$, $\alpha=10^\circ$ and $L_j=3R_c$
is shown in Figure 2. The projection on the $xy$-plane of the
3D outflow cone structure is shown in green. This experiment has
produced $N_{col}=3$ interaction regions (i.e., regions of cone superposition),
shown in white in Figure 2.

There is, of course, a larger
number of ``projected crossings'' between the outflow cones, which
do not correspond to physical interactions. Looking in
the green structure of Figure 2, one counts approximately
$N_{proj}=7$ regions
in which the projected cones cross each other (some of these regions
being superpositions of more than 2 cones). A general result in
all the multi-jet experiments that we have run have that
$N_{proj}\sim 0.5\to 2N_j$.
We do not find any clear trends with the opening angle or with
jet length for $R_j/R_c>1$.

The number $N_p$ of collisions counted in a series of experiments
(a single experiment for each choice of $N_j$, $\alpha$ and $L_j/R_c$)
are shown in Figure 3:
\begin{itemize}
\item the top frame shows $N_p$ as
a function of the half-opening angle $\alpha$ of experiments
with $N_j=10$ and $L_j/R_c=1$ (blue crosses) and 3 (red dots). The
results obtained from equation (\ref{nc}) for the appropriate
values of $\alpha$ and $N_j$ are shown with the solid red
(for $L_j/R_c=3$) and blue ($L_j/R_c=1$) curves. The dashed
curves represent the $\pm \sigma$ curves of the theoretical
$N_{col}$ (see equation \ref{nc}),
\item the bottom frame shows $N_p$ as a function of the
  number of outflows $N_j$ for experiments with $\alpha=5^\circ$ and
  and $L_j/R_c=1$ (blue crosses) and 3 (red dots). The red and
  blue curves are the corresponding analytic predictions obtained
  from equation (\ref{nc}).
\end{itemize}

\section{A gasdynamic simuation}

Using the ``yguaz\'u-a'' adaptive grid, gasdynamic code
(\citealt{raga2000}), we have computed a simulation of a
system of 5 initially conical, bipolar outflows. The jets
are identical, with an $\alpha=10^\circ$ opening angle and
a $L_0=4\times 10^{16}$~cm initial length. A constant $n_j=1000$~cm$^{-3}$
density, $T_j=1000$~K temperature and $v_j=100$~km~s$^{-1}$
radially directed velocity is re-imposed within the
inner part of the cones (of length $L_0/2$)
after each computational timestep. The environment
is initially at rest, with a homogeneous $n_a=1$~cm$^{-3}$
density and $T_a=1000$~K temperature.

For the jet source positions and outflow directions, we have
chosen the randomly sampled position/direction combinations
of the $N_j=5$ model shown in Figure 2. For the gasdynamic
simulation, we have assumed that the source positions are
sampled within a $R_c=5\times 10^{16}$~cm sphere, in the
center of the computational domain.

The simulation is done in a Cartesian grid with an
extent of $2\times 10^{17}$~cm along the 3 axes, using a
4-level binary adaptive grid with a maximum resolution of 
$\approx 3.9\times 10^{14}$~cm (corresponding to
$512^3$ cells at this resolution). Outflow conditions
are applied to all of the grid boundaries.

The gas is assumed to be initially neutral (in the jet cones
and in the environment), except for a small seed ionization
fraction that feeds the collisional ionization cascade. A
rate equation for neutral Hydrogen is solved together with
the gasdynamic equations, and the parametrized cooling function
of \citet{raga2002} is included in the energy equation.

Figure 4 shows a time-series of column density maps (integrated
along the $z$-axis) obtained from our simulations. The initial
outflow cones (top frame) expand almost freely in the low density
environment (other two frames of Figure 4), and produce many lines
of sight superpositions and two seen jet interaction regions.

At $t=400$~yr, the jet cones have reached a length of
$\approx 1.6\times 10^{17}$~cm $\approx 3R_c$ (where $R_c$ is the outer
the radius of the sphere containing the outflow sources, see above).
Therefore, the bottom frame of Figure 4 can be directly compared
with the cone pattern of the geometric outflow model of Figure 2
(noting that Figure 2 covers twice the physical domain of Figure 4).

From this comparison between Figures 2 and 4, it is clear that
the geometric and gasdynamic models share the same distribution
of outflow orientations and source positions. Looking at the
geometric model of Figure 2, we see that this configuration of
outflows have 3 superposition regions (in white):
\begin{itemize}
\item The upper one is seen as a major jet collision in
  the numerical simulation, resulting in a dense, approximately
  interaction region in which the two colliding
  cones are redirected in an approximately vertical direction
  (see the bottom frame of Figure 4),
\item the lower cone superposition region also appears as a dense
  two-jet interaction region in which part of the interacting
  cones are redirected to the bottom right direction,
\item the central, small cone superposition region (see Figure 2)
  can be seen as a column density peak (in the bottom frame of Figure 4),
  but does not produce a major redirection of the impinging jets.
\end{itemize}
Therefore, two out of the three cone superposition regions are seen
in the numerical simulation, there are major disruptions in the colliding outflows.
The third cone superposition of our distribution of outflows is only
a grazing collision involving only a part of the cross-sections
of the colliding jets.

\section{Application to observed outflow systems}

Let us assume that we have an image of a system of $N_j$ bipolar
outflows. We can then calculate:
\begin{itemize}
\item the mean tangent of the half-opening angle $\overline{\tan \alpha_p}$
  of the projected cones,
\item the mean length $\overline{L}_p$
of the jets (measured from the sources to the tip of one of the
outflow lobe),
\item the root mean square separation $\overline{\Delta R}$
  between all the possible source pairs.
\end{itemize}
If we assume that the jets have the same intrinsic length,
for a random direction distribution of the outflows, we
can estimate the true (un-projected) half-opening angle $\alpha$ of
the outflows through:
\begin{equation}
  \tan \alpha=0.59\,\overline{\tan \alpha_p}\,,
  \label{tana}
\end{equation}
and the true (un-projected) length of the jets as:
\begin{equation}
  L_j=1.70\,\overline{L}_p\,.
  \label{lj}
\end{equation}
Also, if we assume that the source positions are randomly
distributed within a spherical volume of radius $R_c$. we
can estimate this radius through:
\begin{equation}
  R_c=0.92\,\overline{\Delta R}\,.
  \label{rc}
\end{equation}

As an illustration of how this method can be applied to an
observed system of outflows, we apply it to the NGC~1333 region.
\citep{d08} showed that this region has a number of
superimposed outflows (see also \citealt{raga2013}). If we consider
the outflows from the YSO\,15, 18, 20, 22 23 and 32 sources (with
2 outflows arising from the YSO\,23 source), we have a system
of $N_j=7$ outflows. Considering a distance of 220~pc to NGC~1333,
from Figure 7 of \citep{d08}, we find a $\overline{\Delta R}
=6.96\times 10^{17}$~cm mean square separation between the outflow
sources and a mean projected length $\overline{L}_p=1.48\times 10^{18}$~cm
for each outflow lobe.

Using equations (\ref{lj}-\ref{rc}) we obtain an estimated
de-projected jet length $L_j=1.36\times 10^{18}$~cm and a radius
$R_c=1.18\times 10^{17}$~cm, therefore obtaining $L_j/R_c\approx 0.87$.
Now, $N_j=7$ and $L_j/R_c\approx 0.87$, from equation (\ref{nc}),
we find:
\begin{equation}
  N_{col}\approx 1.1(\alpha/10^\circ)^{1.3}/2\,.
  \label{nc1333}
\end{equation}

Now, it is not straightforward to estimate the half-opening angle
$\alpha$ of the NGC~1333 outflows. From the Spitzer image in
Figure ~4 of \citet{raga2013}, we estimate a mean half-opening
angle $\alpha_p\approx 5^\circ$, which through equation (\ref{tana})
gives an deprojected $\alpha=3^\circ$ angle. Inserting this angle
in equation (\ref{nc1333}), we obtain $N_c\approx 0.11$, with a standard
deviation of 0.34.

Therefore, for the NGC~1333 system, the expected number of physical
collisions between the observed outflows is zero. This 
is only the statistically expected value, and there is a non-zero
probability for actual collisions taking place.

An estimate for the probability of a collision between
two of the jets in NGC~1333 can be estimated as follows. Let
us consider the jets from YSO23 and YSO24
of \citep{d08}. The two outflow
sources have a projected separation of $\sim 15''$, and
their projected lengths are considerably longer
If we use the projected separation and length as
estimates of the deprojected separation and length,
we conclude that these two jets are in the ``large $L_j/L_{12}$''
regime of a two-jet interaction (see Figure 1). From equation
(\ref{p}), for $\alpha=3^\circ$ and $L_j/L_{12}\gg 1$ we
obtain $P_{col}\approx 0.011$. This estimate tells us that there
is a low probability of a few times $\sim 1$\,\% of
having a jet collision between two given jets.

To complement our analysis, we now examine a more massive and significantly denser star-forming environment: the Orion Nebula Cluster (ONC) (\citealt{Kroupa2018}). This region is known to host a significant number of young stellar objects \citep{alves2012orion}, most of which are likely to drive bipolar, collimated outflows. By applying the same statistical framework to Orion, we aim to investigate the dynamic consequences and enhanced interaction rates resulting from the significantly higher stellar density.

We assume a system composed of $N_j = 10^4$ randomly distributed bipolar outflows, corresponding to the approximate number of young stars in the Orion cluster \citep{megeath2012spitzer}. We adopt a typical intrinsic jet length of $L_j = 0.1\,\mathrm{pc}$ and consider a spatial distribution confined within a spherical region of radius $R_c = 2\,\mathrm{pc}$, roughly corresponding to the size of the central ONC core \citep{Kroupa2018}. For the intrinsic half-opening angle of the jets, we assume a value of $\alpha = 3^\circ$, consistent with the observational constraints from collimated jets in young stars \citep{erkal2021}. Using the full analytical formulation for the expected number of collisions (equation \ref{nc}), we find a collision probability of $P_{col}\sim 0.65\%$. Although this jet-jet collision probability is low, each individual jet is expected to experience approximately 66 interactions on average. This outcome illustrates the combinatorial amplification effect: even a small probability, when considered over a large ensemble of jets, results in a substantial cumulative effect on the dynamical evolution of the cluster.




We can also increase the half-opening angle to $\alpha = 10^\circ$, while keeping $L_j/R_c = 0.05$, which results in a substantial rise in the expected probability of collision: $P_{col}\sim 3.14\%$, leading to an average of 314 collisions per jet. This parameter variation demonstrate that, although changes in jet length and cluster size moderately affect interaction rates, variations in the angular width have the most dominant effect, dramatically enhancing the probability of jet-jet interactions in dense star-forming environments.

In the case of Orion, we have chosen to focus on reporting the average number of collisions per jet rather than the total number of collisions. This choice is motivated by the fact that our analytical framework does not impose a limit on jets that have already interacted, implicitly treating each possible pairwise interaction as independent, even after multiple encounters. In other words, the model assumes that once a jet has interacted, it remains geometrically and dynamically available for further interactions as if no physical alteration occurred. As discussed in Section 3, when the total number of collisions becomes comparable to the total number of jets, a more complex hypergeometric distribution is required to capture the statistics accurately. We do not explore this limit here, as a real system of outflows with many mutual interactions would defy our simplified picture of geometrically fixed biconical structures.


\subsection{Volume filling factor}

An important consideration that emerges from this statistical framework is the connection between the number of outflow interactions and the fraction of the volume evacuated by jets. Since each bipolar outflow traces a biconical volume, the cumulative effect of many such jets is to fill a significant portion of the host cluster's volume. Once a critical fraction is filled, geometric overlap between outflows becomes statistically inevitable. A useful quantity to describe this collective effect is the volume filling factor, defined as the fraction of the cluster volume dynamically influenced or occupied by the jets at a given time.

The volume of a single bipolar outflow is approximated as the sum of two identical cones. The instantaneous volume cleared by a single bipolar jet is thus given by:
\begin{equation}
    V_{jet}(t) = 2 \times \frac{1}{3} \pi L_j(t)^3 \tan^2{\alpha}
\end{equation}
 where $L_j(t)$ is the instantaneous jet length, and $\alpha$ is the intrinsic half-opening angle of each cone. When considering a total of $N_j$ jets, randomly distributed within a spherical region of radius $R_c$, the total evacuated volume becomes:
\begin{equation}
    V_{tot}(t) = N_j \times V_{jet}(t). 
\end{equation}

We then define the volume filling factor as:
\begin{equation}
    f_V(t) = \frac{V_{tot}(t)}{V_c},
\end{equation}
where $V_c = \frac{4}{3}\pi R_c^3$ denotes the total volume of the cluster region under consideration. In this case, the volume filling factor simplifies to:
\begin{equation}
   f_V(t) = \frac{N_j L_j(t)^3 \tan^2\alpha}{2 R_c^3}.
\end{equation}
This expression reveals that the volume filling factor depends cubically on the jet length and quadratically on the opening angle, thereby embedding direct information about both jet dynamics and geometry. As jets grow over time during YSO evolution phase, the volume filling factor tends to increase, indicating the progressive occupation and dynamical influence of the outflows within the cluster volume. Physically, $f_V(t)$ quantifies the fraction of the volume of the cluster that is dynamically influenced or evacuated by the jets at a given time. When the volume filling factor is small, jets remain mostly isolated and interactions are rare; as approaches unity, the system transitions into a highly interactive regime where overlapping cavities and mutual collisions become increasingly probable.

\begin{figure}[!t]
    \includegraphics[width=0.95 \columnwidth]{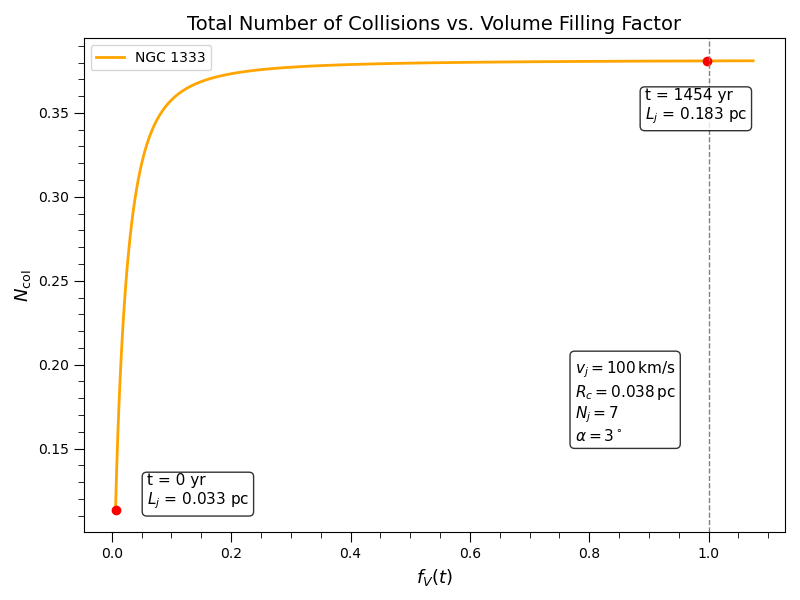}
    \caption{Total expected number of jet-jet collisions, $N_{col}$, as function of the volume filling factor, $f_V(t)$, for the NGC 1333 region. The curve corresponds to a system of $N_j =7$ bipolar outflows, each with an initial jet length of $0.033\,\mathrm{pc}$, confined within a spherical cluster of radius $R_c = 0.038\,\mathrm{pc}$, and adopting a half-opening angle of $\alpha = 3^\circ$. The initial and final points are highlighted in red. The dashed vertical line marks the point where the volume filling factor reaches unity ($f_V = 1$), indicating that jets fully occupy the cluster volume. Text boxes indicate key physical parameters, including the jet velocity, characteristic scales, and corresponding jet lengths at initial and final times. This graph illustrates how collision rates increase as the jets dynamically expand and progressively fill the available volume.}
    \label{fig5}
\end{figure}

\section{Conclusions}

A simple model for interaction in a system of
identical conical bipolar outflows is presented. We first compute (with
a Montecarlo approach) the probability of having a collision between
two outflows.

As expected, the probability $P_{col}$ of
having an interaction (which corresponds to a partial superposition
of the outflow cones in our simple geometric model) is
a function of their half-opening angle $\alpha$ and of the
ratio $L_j/L_{12}$ between the length of the outflow lobes and
the separation between the outflow sources. As expected, we find
that $P_{col}=0$ for $L_j/L_{12}<1/2$ and also a $P_{col}\to P_\infty(\alpha)$
for $L_j/L_{12}\gg 1$. This behaviour is shown in Figure 1.
We find an analytic fit for $P_{col}(L_j/L_{12},\alpha)$ (see Section 2
and Figure 1).

Building upon this, the derived interaction probability for two-outflows is used to
estimate the expected number of jet interactions in a system composed of $N_j>2$ identical outflows (see Section 3).
Assuming that the interactions follow
a binomial distribution (as appropriate for a system with only
a few jet interactions), we then derive an analytic expression
(equation \ref{nc}) for the expected number of jet collisions
for a system of bipolar jets with random directions ejected from
$N_j$ sources that are uniformly distributed within a spherical
``cluster'' of radius $R_c$. The result is a simple analytic expression
giving the expected number $N_c$ of two-jet collisions as a function
of the half-opening angle $\alpha$ of the outflow cones and the
jet length to cluster radius ratio $L_j/R_c$ (see equation \ref{nc}).

Additionally, a series of Montecarlo simulations are carried out for systems of $N_j=5\to 25$ identical outflows with half-opening angles
$\alpha=2\to 15^\circ$ and lengths $L_j=(1,\,3)\times R_c$, with which
we check our analytic estimate for the expected number $N_c$ of
outflow collisions which we derived analytically (equation
\ref{nc}). A satisfactory agreement is obtained between
the Montecarlo simulations and the analytical
$N_c(L_j/R_c,\alpha)$ (see Figure 3).

In order to explore the properties of systems of interacting
jets beyond our simple, biconical outflow geometric model, we have
carried out a single, 3D gasdynamical simulation of a system of 5
initially biconical outflows. For the chosen configuration of source
positions and outflow directions, the geometric model gives 3 ``cone
superposition'' regions (Figure 2). Two of these superposition
regions are seen as major two-jet interactions in the gadynamical simulation,
with high compressions and substantial redirection of the interacting
outflows (see Figure 4). The third cone superposition region (of the
geometric model, see Figure 2) is seen as a minor, grazing collision
in the gasdynamical model (see Figure 4).

Finally, we describe how to calculate the expected number $N_c$
of two-jet collisions in a system of $N_j$ outflows. We propose
calculating the root mean square of the projected separations
between the outflow sources to estimate the physical radius $R_c$
of the region within which the sources are distributed. Also, the
mean projected length of the jets can be used to estimate the
(deprojected) average jet length. Finally, the half-opening
angle $\alpha$ has to be estimated from the observations
(deprojecting the observed, projected opening angles of the outflows).
With the resulting observational determinations of $N_j$, $\alpha$ and
$L_j/R_c$, one can use equation (\ref{nc}) to calculate the expected
number $N_c$ of two-jet collisions in the observed outflow system.
We have illustrated this procedure using published observations
of the NGC~1333 outflow system.

When trying to obtain the parameters necessary for
calculating $N_c$ (through equation \ref{nc}), the half-opening angle $\alpha$ is the most uncertain. It is unclear where
the edges of the outflow lobes are and whether one has to consider
a more central, jet-like component or the broader envelope
seen in some outflows. Given this difficulty, it is 
possible to appropriately invert equation (\ref{nc}) to obtain
the opening $\alpha$ angles necessary for obtaining $N_c\geq 1$.

In this paper, we have derived a relatively simple way
to estimate the expected number of two-jet collisions in a system
with many outflows. Its application to many observations is 
possible.

It is important emphasizes that the statistical estimates derived here rely on the assumption that the outflow sources are randomly distributed throughout a cluster volume and that the jet orientations are isotropic. However, in real star-forming regions, the distribution of young stellar objects is often far from uniform. Stellar clustering, filamentary gas structures, and the high prevalence of binary and multiple systems introduce significant spatial correlations among sources. These effects are expected to increase the probability of jet-jet interactions, particularly in compact binaries or small groups where projected separations are small. Consequently, the number of physical overlaps estimated under the uniform distribution assumption should be considered as a lower limit. More realistic models accounting for stellar multiplicity and substructure may yield higher collision probabilities.

\acknowledgments
Dr. Alejandro Raga, lead author of this work, passed away on July 20, 2023. A couple of weeks prior, the co-authors provided the final comments to update the manuscript before submission. The authors want to thank Pablo Vela\'{z}quez and Ary Rodr\'{ı}guez (Instituto de Ciencias Nucleares) for their help in recovering Dr. Raga’s files related to this work. Dr. Raga was an amazing friend and colleague. He will be fondly remembered and sadly missed.\\

This work was supported by the DGAPA (UNAM) grant IG100422 and IN110722.

\vfill\eject
\bibliography{reference-file.bib} 
\vfill\eject

\end{document}